\documentclass[debug,overfull]{epl_avec_numero_de_preprint}
\input{psfig.sty}
\usepackage{amsfonts}
\usepackage{amsmath}
\usepackage{graphicx}
\usepackage{xspace}

\newcommand{\id}{1\hspace{-2.6pt}\mathrm{I}}
\newcommand{\ie}{\textit{i.e.}\ }
\newcommand{\eg}{\textit{e.g.}\ }
\newcommand{\dd}{\mathrm{d}}
\newcommand{\Psuccess}{P_\mathrm{succ}}
\newcommand{\alphaA}{\alpha_\mathrm{A}}
\newcommand{\alphaflat}{\alpha_\mathrm{R}}
\newcommand{\exptemps}{\tau}
\newcommand{\expcc}{\gamma}
\newcommand{\expproba}{\lambda}
\newcommand{\exppzero}{{\gamma _0}}
\newcommand{\expepsilon}{\theta}
\newcommand{\Ai}{\mathrm{Ai}}
\newcommand{\adeux}{\delta_2}
\newcommand{\citefootnote}[1]{\ensuremath{(^#1)}}

\title{Critical behaviour of combinatorial search algorithms, and 
the unitary-propagation universality class}
\shorttitle{Critical behaviour of search algorithms}

\author{C. Deroulers \and R. Monasson}
\institute{CNRS-Laboratoire de Physique Th{\'e}orique de l'ENS,
24 rue Lhomond, 75005 Paris, France.}

\begin{document}
\maketitle

\begin{abstract}
The probability $P(\alpha, N)$ that search algorithms for random
Satisfiability problems successfully find a solution is studied as a
function of the ratio $\alpha$ of constraints per variable and the number
$N$ of variables. $P$ is shown to be finite
if $\alpha$ lies below an algorithm-dependent threshold $\alphaA$, 
and exponentially small in $N$ above.
The critical behaviour is universal for all algorithms based on the
widely-used unitary propagation rule: 
$P[ (1 + \epsilon)\, \alphaA , N]
\sim \exp (- \, N^{1/6} \, \Phi (\epsilon N^{1/3}))$. 
Exponents are related to the critical behaviour of random
graphs, and the scaling function $\Phi$ is exactly calculated through a
mapping onto a diffusion-and-death problem. 
\end{abstract}

\preprint{LPTENS 04/25}

{\em Introduction}.
The discovery of universality in phase transition phenomena was a
major progress in modern condensed matter and statistical physics. The
purpose of this letter is to point out that universality also takes
place in computer science, more precisely, in
computational complexity theory. There, the goal is to understand
whether a computational task consisting in processing a large number
$N$ of input data can be carried out in a time scaling only polynomially \eg
$N^3$, and not exponentially \eg $2^N$, with $N$ \cite{papa}. Depending on
input data defining parameters, dynamical phase transitions between these
two behaviours may take place \cite{kirk,extra2,chap}. We prove hereafter, 
for the case of the celebrated Satisfiability (SAT) problem \cite{papa}, 
that the onset of complexity at criticality is
universal in that it depends on some structural features of resolution
algorithms and input data statistics only.

{\em Definitions of computational task and algorithm.}
In the random K-SAT problem \cite{kirk}, one wants to
find a solution to a set of $M=\alpha N$  randomly drawn
constraints (clauses) over a set of $N$ Boolean variables $x_i$ 
($i=1\ldots N$). Each constraint reads $z_{i_1} \vee z_{i_2} \vee
\ldots \vee z_{i_K}$, where $\vee$ denotes the logical OR; $z_\ell$ is a
variable $x_{i_\ell}$ or its negation $\bar x_{i_\ell}$ with equal
probabilities ($=\frac 12$), and $(i_1, i_2, \ldots , i_K)$ is a $K$-uplet
of distinct integers unbiasedly drawn from the set of the $\binom{N}{K}$
$K$-uplets. We now study the $K=3$ case, the smallest value
for which the problem is NP-complete \cite{papa}, and 
$K\ge 4$ later.

Our algorithms start from {\em tabula rasa}, then iteratively assign
variables to true ($T$) or false ($F$) according to two well-defined rules
(specified below), and modify the constraints accordingly \eg $\bar
x_1\vee \bar x_2 \vee x_3$ becomes $\bar x_1\vee x_3$ if $x_2=T$
\cite{kirk,Fri}. At the end, no constraint is left (a solution is found
--- success case), or a contradiction is found (one variable previously
assigned to, say, $T$ is required to be $F$ from modified constraints ---
failure case). The first assignment rule, UP (for unitary propagation)
\cite{Fri}, is common to all algorithms: if a clause with a unique
variable is produced at some stage of the procedure \eg $\bar x_1$, then
this variable is assigned to satisfy the clause \eg $x_1=F$. The second
rule is a specific and arbitrary prescription taking over UP when it
cannot be used \ie in the absence of unique variable clause. In the
simplest algorithm, referred to as R (random), the prescription consists
in setting any unknown variable to $T$ or $F$ with prob. $\frac 12$
independently of the remaining clauses \cite{Fri}; more sophisticated
prescriptions \cite{kkl,dub} will be studied later.

Resolution procedures used in practical applications are based on the 
combination of the above algorithm and a backtracking principle 
\cite{kirk}: in case of failure, the last variable assigned
through the prescription (not through UP) is flipped, and the algorithm
resumes from this stage. At the end, either a solution is found or all
possible backtracks have failed, and a proof of the absence of solution is
obtained. The resolution time typically scales as $O(N)$ if $\alpha < 
\alphaA$ and $\exp O(N)$ if $\alpha >\alphaA$, where the threshold 
$\alphaA$ depends on the algorithm. Intuition suggests and
analyses prove \cite{chap,rigtrans} that this poly/exp crossover 
is due to the success/failure transition of the pure algorithm
\ie without backtracking. More precisely, $\alphaA$ can
be identified with the ratio at which the probability
$\Psuccess(\alpha , N)$ of success of the pure algorithm 
vanishes as $N\to \infty$ \cite{Fri}. 
To understand the onset of complexity at $\alphaA$, it is thus
natural to analyze how $\Psuccess$ vanishes when the ratio $\alpha$ is
kept close to its critical value and $N$ increases.

\begin{figure}
\begin{center}
\includegraphics[width=230pt,height=200pt]{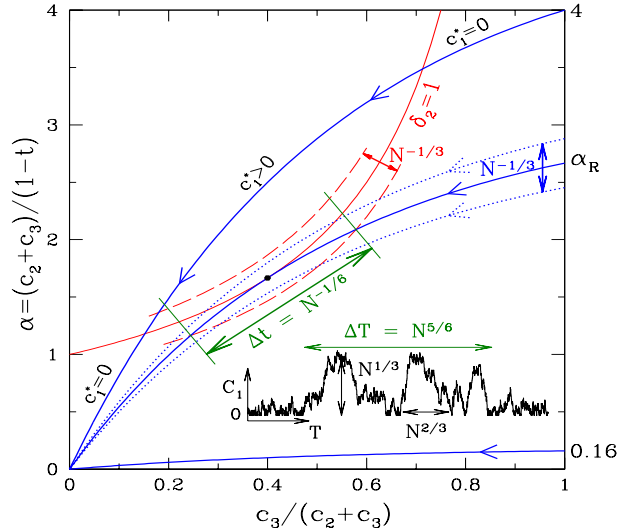}
\caption{\label{figure_traj} Resolution trajectories for the R
algorithm  (bold lines, arrows indicate
time direction, from a fraction $t=0$ of eliminated variables 
--- right axis --- to $t=1$ --- lower
left corner where a solution is found). 
For initial ratio  $\alpha < \alphaflat$, $C_1$ stays bounded 
(success case). When $\alpha > \alphaflat$, $C_1 \sim N$ when the 
trajectory lies above the contradiction line $\adeux=1$, and the
density $c_1^*$ is positive (failure case). At the critical
ratio $\alphaflat (=\frac 83)$, the trajectory hits tangentially
(black dot) the contradiction line. The critical region is defined by
fluctuations $\sim N^{-1/3}$ for finite size $N$ around these two
lines (dotted and dashed lines respectively), and is crossed 
through assignment of a fraction $\Delta t= N^{-1/6}$ of variables.
\textbf{Inset:} $C_1$ vs. $T$  in the critical region for a particular  
run with $N=10^5$ and $\alpha = \alphaflat$. Reported scalings 
correspond to the largest components ($S\sim N^{2/3}$).}
\end{center}
\vskip -.8cm
\end{figure}

{\em Analysis of the R algorithm}.
Each time a new variable is assigned some clauses are eliminated, other
are reduced or left unchanged. We thus characterize the set of clauses
by its state $(C_1 (T),C_2(T),C_3(T))$, where $C_j$ is the number of
$j$-clauses \ie involving $j$ variables ($j=1,2,3$) and $T$
is the number of assigned variables \cite{Fri,chap}. 
Consider a 3-clause left at `time' $T$. When $T\to T+1$, the newly
assigned variable has a probability $3/(N-T)$ to appear (as is, or
negated) in this
3-clause; if so the clause will be satisfied or reduced into a 2-clause
(with equal prob. $\frac 12$). As a consequence the average change of $C_3$
equals $-3 C_3 (T) / (N-T)$. In the large $N$
limit, the density $c_3(t)= C_3(t\,N)/N$ of $3$-clauses becomes
concentrated around its average value, solution of the ordinary
differential equation $\dd c_3/\dd t = -3 c_3/(1-t)$. A similar reasoning
leads to $\dd c_2/\dd t = 3 c_3/2/(1-t) - 2 c_2 /(1-t)$ for the density 
$c_2$ of 2-clauses. Solving these
equations with initial conditions $c_3(0)=\alpha, c_2(0)=0$ gives
$c_3(t)=\alpha (1-t)^3, c_2(t) = \frac 32 \alpha t(1-t)^2$ and 
the resolution trajectories of Fig.~\ref{figure_traj} \cite{chap}.

The above evolution for $c_2,c_3$ is correct as long as no
contradiction has emerged as a result of the production of two opposite 
1-clauses \eg $x_1$ and $\bar x_1$. 
The probability $P_N(C_1; T)$ that the
assignment of $T$ variables has produced no contradiction and a set of
constraints with $C_1$ 1-clauses obeys a Markovian
evolution from $T$ to $T+1$ with a transition matrix \cite{chap},
\begin{eqnarray}
\label{bbra}
&& H_{N} [ C'_1 \leftarrow C_1;T, C_2] =
\sum_{s_2,r_2} {\cal M}_{p_2}^{C_2; s_2,r_2} \Big[
\id_{C_1} \id_{C'_1-r_2} \nonumber \\ &&
+ (1-\id_{C_1}) \sum_{s_1} {\cal M}_{p_1}^{C_1-1; s_1, 0}
\id _{C'_1-C_1 + 1 + s_1 - r_2} \Big]
\end{eqnarray}
where $\id_C$ denotes the Kronecker function: $\id _C \equiv 1$
if $C=0$, $0$ otherwise. Variables appearing in (\ref{bbra}) are as
follows: $s_j$ (respectively $r_j$) is the number of $j$-clauses
which are satisfied (resp. reduced to $j-1$ clauses) when the
$(T+1)^{th}$ variable is assigned. These are stochastic variables drawn
from multinomial distributions ${\cal M}_{p} ^{C;x,y}\equiv \binom{C}{x,y}
p^{x+y} (1-2 p)^{C-x-y}$.  
Parameter $p_j \equiv j/2/(N-T)$ equals the probability
that a $j$-clause contains the variable just assigned;
$r_1=0$ demands that no opposite 1-clauses and thus no contradiction
are present. Equation (\ref{bbra}) defines a random motion for a walker 
moving on the semi-infinite line $C_1\ge 0$ with time-dependent and 
random (through $C_2$) rates. The success/failure
transition takes place when the average number $\adeux =
c_2/(1-t)$ of 1-clauses created from 2-clauses (right move) exceeds 
the number of 1-clauses ($1$ \footnote{More than one clause
can be satisfied especially when $C_1\sim N$ \ie in the failure regime.}) 
eliminated by UP each time a variable is assigned (left move) \cite{Fri}. 

\begin{figure}
\begin{center}
\includegraphics[width=230pt,height=200pt]{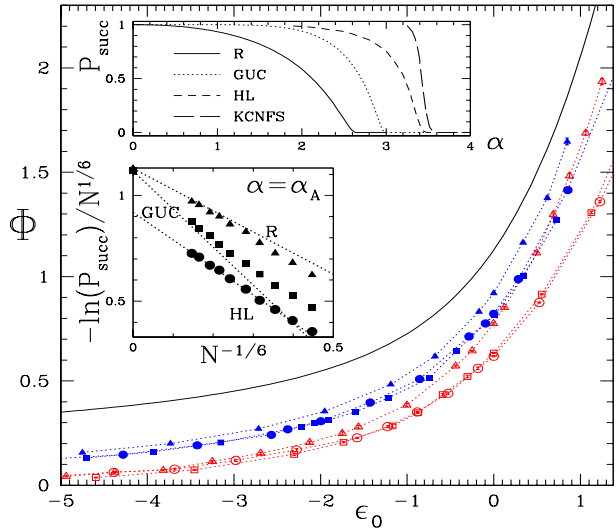}
\caption{\label{figure_Phi} Scaling function $\Phi$ (solid line) 
compared to numerical simulations for $N=1000$
(empty), $20000$ (filled symbols) and algorithms
R ($\triangle$), GUC ($\square$) and HL ($\bigcirc$). 
Error bars (with $\simeq 10^5$ samples) 
are smaller than symbol size. Data for GUC and HL are 
rescaled horizontally and vertically, see text. Dotted lines serve as a
guide for the eye. \textbf{Bottom inset:} $-\ln(\Psuccess)/N^{1/6}$ vs. 
$N^{-1/6}$ at the critical thresholds $\alpha_\mathrm{R}, 
\alpha_\mathrm{GUC} \simeq 3.003, \alpha_\mathrm{HL} \simeq 3.425$. Linear 
fits (dotted lines) extrapolate to theoretical predictions for $\Phi(0)$ 
(available for R and GUC) on the left axis. \textbf{Top inset:} 
$\Psuccess$ vs. $\alpha$ showing the algorithm-dependent success/failure 
transition. Curves are analytical for R, GUC, and numerical for HL, 
KCNFS.}
\end{center}
\vskip -.8cm
\end{figure}

{\em Successful regime ($\adeux<1$).} 
If elimination of 1-clauses is faster than creation, $C_1$ stays
bounded throughout the search process. On time scales $1\ll T\ll
N$, $C_1$ reaches equilibrium, with a distribution $p_0 (C_1,t)$
function of slow variables only \ie $c_2,t$.
This, and the probability of success can be derived
with the simple Ansatz $P_N(C_1;T) = p_0(C_1,t) + p_1(C_1,t)/N
+O(N^{-2})$ and sending $N\to\infty$. We find that $p_0(C_1,t) \sim
\exp (-\rho \,C_1)$ at large $C_1$, where $\rho$ is the time-dependent 
positive root of $\rho / \adeux(t) =1-e^{-\rho}$. As expected,
$\rho>0$ and $C_1$ is bounded as long as $\adeux <1$. At fixed ratio 
$\alpha$, $\adeux$ reaches its maximum $\adeux^M = \frac 38 \alpha$ 
along the resolution trajectory
for a fraction $t_R = \frac 12$ of assigned variables; 
the transition takes thus place at $\alphaflat = \frac 83$
\cite{Fri}.  
The probability of success is given by $\Psuccess
= \sum _{C_1} p_0(C_1,1) = \exp [ \frac 1{2r} - \hbox {\rm arctan}
(1/\sqrt{r-1}) / 2/\sqrt{r-1} ]$ with $r= \alphaflat/\alpha$, and is 
shown in Fig.~\ref{figure_Phi} (top inset). Note that $- \ln 
\Psuccess [\alphaflat (1-\epsilon)] \sim \frac{\pi}4\, 
{\epsilon}^{-1/2}$ as $\alpha$ reaches
its critical value $\alphaflat$ from below \cite{Moore}.

{\em Failure regime ($\adeux>1$).} 
When $\alpha > \alphaflat$, the resolution
trajectory crosses the contradiction line $\adeux=1$ 
(Fig.~\ref{figure_traj}). 
$C_1$ then becomes of the order of $N$ and each assignment has a finite 
probability to produce some contradiction; the probability of success 
is thus exponentially small with $N$ \cite{amai}. 
Later the trajectory crosses the contradiction line back
and $C_1=O(1)$ again. This behaviour is captured with 
the Ansatz $P_N(C_1;T) = \exp(- N
\omega (c_1,t))$ where $\omega$ is the rate function associated to
the large deviations of the density $c_1$ of 1-clauses. $\omega$ fulfills a 
first order partial differential equation  \cite{chap,amai},
which can be explicitly solved for ratios slightly above the threshold \ie 
$\alpha = \alphaflat(1+\epsilon)$. The
coordinates $c_1^*, \omega^*$ of the minimum of $\omega$ \ie
the most likely trajectory are, at time
$t= \frac 12 (1+ s \sqrt \epsilon)$: $c^*_1=0, \omega^*=0$ if $s < -1$; 
$c_1 ^*= \frac 12 (1-s^2)^2 \epsilon ^2$, $\omega ^*= [-
\frac {s^5}{20} - \frac  {s^3}{6} + \frac{s}4 +\frac 2{15}]
\epsilon ^{5/2}$ if $|s | < 1$; 
$c^*_1=0, \omega^*=  \frac 4{15} \epsilon ^{5/2}$ if $s > 1$.
Thus, $-\ln \Psuccess \sim \frac 4{15} \epsilon ^{5/2}N$ to the lowest 
order in $\epsilon$.  


{\em Critical regime ($\adeux\simeq 1$)}.
For $N$ large but finite and $\alpha$ close to
$\alphaflat$, finite-size scaling \cite{kirk} applies if 
\begin{equation} \label{fss}
- \ln \Psuccess  \bigg((1+\epsilon)\,  \alphaflat , N \bigg)  
= N^\expproba \ \Phi \big( \epsilon\; N^\expepsilon \big)
\end{equation}
for some regular function $\Phi$. In the infinite size $N$ limit,
this expression should agree with the above results
for the successful ($\epsilon<0$) and failure ($\epsilon>0$) regimes.  
Matching the powers in $N$ and $\epsilon$, we find $\expproba - \frac
{\expepsilon}2=0$, 
$\Phi(x) \sim \frac{\pi}4\, |x|^{-1/2}$ as $x \to -\infty$, 
and $\expproba + \frac{5 \expepsilon}2 = 1$, 
$\Phi(x) \sim  \frac{4}{15}\, x^{5/2}$ as
$x \to +\infty$.  As a result, $\expproba=\frac 16$ and 
$\expepsilon= \frac 1 3$. Figure~\ref{figure_Phi} (lower inset) shows the
good agreement of numerical experiments performed at the critical
point with the prediction $- \ln \Psuccess \sim
N^{1/6}$.

Introduction of the oriented graph $G$ representing 1- and 2-clauses
allows us to understand the above scalings. 
$G$ is made of $2\, (N-T)$ vertices
(one for each variable $x_i$ and its negation $\bar x_i$), $C_1$
marked vertices (one for each 1-clause $z_i$), and $2\, C_2$
edges ($z_i\vee z_j$ is represented by two oriented edges 
$\bar z_i \to z_j$ and $\bar z_j \to z_i$) \cite{chayesetal}.
$\adeux$ is simply the average (outgoing) degree of vertices in $G$.
A step of UP corresponds to removing 
a marked vertex (and its attached outgoing edges), after having marked its 
descendents; steps are repeated until the connected component is
entirely removed (no vertex is marked). 
Then a vertex is picked up according to the prescription and marked,
and UP resumes. A contradiction arises when two `opposite' vertices \ie 
associated to opposite variables are marked. The 
success/failure transition coincides with the percolation 
transition on $G$ \ie $\adeux=1$ as expected.
From random graph theory \cite{luczak,chayesetal}, in the
percolation critical window $|\adeux -1| \sim N^{-1/3}$, the 
probability that a vertex belongs to a component of size $S$ is 
$Q(S)\sim S^{-3/2}$, with a cut-off equal to the largest size,
$N^{2/3}$ \cite{rgscaling}. 
From  Fig.~\ref{figure_traj}, departure ratios $\alpha$ have to differ
from $\alphaflat$ by $N^{1/3}$ for resolution trajectories
to fall into the critical window. Hence $\expepsilon= \frac 13$.
The time spent by resolution trajectories in the critical
window is $\Delta t \sim \sqrt {|\adeux -1|} \sim N^{-1/6}$,
corresponding to $\Delta T = N \, \Delta t \sim N^{5/6}$ eliminated
variables. Let $S_1, S_2, \ldots , S_J$ be the
sizes of components eliminated by UP in the critical window; 
we have $J\sim \Delta T/ \int \! \dd S \, Q(S) \, S 
\sim N^{1/2}$. During the $j^{th}$ elimination, the
number of marked vertices `freely' diffuses, and reaches $C_1\sim \sqrt
{S_j}$ (Inset of Fig.~\ref{figure_traj}). The probability
that no contradiction occurs
is $[(1-q)^{C_1}]^{S_j} \sim e^{- S_j^{3/2}/N}$ where $q\sim
\frac 1N$ is the probability that a marked vertex is `opposite' to
the one eliminated by UP.
Thus $-\ln \Psuccess \sim J \! \int \! \dd S \, Q(S) \, S^{3/2}/N \sim 
N^{1/6}$, giving $\expproba = \frac 16$. Notice that, while the average 
component size is $S\sim N^{1/3}$ (and thus $P_N(C_1=0)\sim N^{-1/3}$), 
the value of $\expproba$ is due to the largest components with $S\sim 
N^{2/3}$ \ie $C_1 \sim N^{1/3}$ marked vertices. 
The distribution of $c=C_1/N^{1/3}$ is calculated below.

\begin{figure}
\begin{center}
\includegraphics[width=140pt,angle=-90]{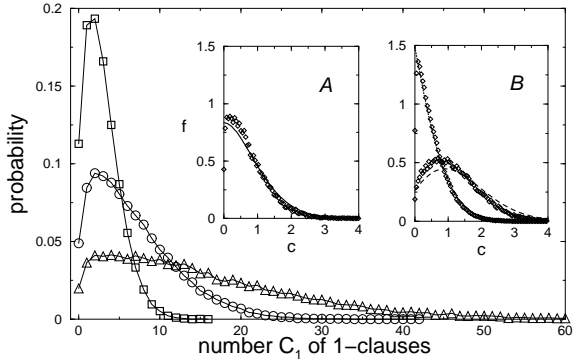}
\caption{\label{trace_rho} Histograms of the numbers $C_1$ of 1-clauses 
for sizes $N=10^2$ ($\square$), $10^3$ ($\bigcirc$) and $10^4$ 
($\triangle$) right at criticality 
(time $t=\frac 12$ and ratio $\alpha=\frac 83$). Note the sharp
increase of the probability around $C_1=0$, in quantitative 
agreement with the theoretical prediction $f_0=f(0)/2$.
\textbf{Insets:} theoretical distributions $f$ for
$c=C_1/N^{1/3}$ at criticality (A, same data as main figure), and
for drifts $v_0=0.5$ (B, dotted),  $v_0=-0.7$ (B, dashed curve) 
compared to numerics.}
\end{center}
\vskip -.8cm
\end{figure}

{\em Scaling function.}
To calculate $\Phi$ in (\ref{fss}),  we magnify the critical region 
in Fig.~\ref{figure_traj} and consider ratios $\alpha = \frac 83 (1 + 
\epsilon _0 \, N^{-\expepsilon})$ and times 
$t= \frac 12 \, (1 + t_0\, N^{-\exptemps})$, where $\expepsilon$ and
$\exptemps$ are scaling exponents to be determined.
We then decompose the probability of having $C_1$ clauses in the
critical region into the
product ${P}_{N}(C_1,T)  = \exp [- N^\expproba \varphi(t_0)]
\times F(C_1,t_0)$; the first term is the probability that
no contradiction has been found up to `time' $t_0$, and $F$ is the
(normalized) probability distribution of 1-clauses.
Clearly $\varphi(t_0 \to -\infty )=0$ since the probability that the
search process has ended is not vanishingly small before the
trajectory enters the critical region (Fig.~\ref{figure_traj}). 
We make the Ansatz $F(C_1) = N^{-\expcc} 
f(c=C_1 N^{-\expcc}) $ where $f$ is the probability distribution 
of the rescaled number $c$ of unit-clauses (Fig.~\ref{trace_rho}).
Last of all, the probability that a variable is set through a free
choice and not UP, $P_N(C_1=0)$, is assumed to scale as $f_0 /N^\exppzero$. 

The evolution equation for $P_N$  based on matrix (\ref{bbra}) 
imposes $\expepsilon=\expcc=\exppzero=\frac 13$, 
$\expproba=\exptemps=\frac 16$ in agreement with the above scaling
arguments \footnote{These results are not affected by the presence of
Gaussian fluctuations $\sim N^{-1/2}$ in $c_2(t)$.}.%
\edef\notegaussienne{\thefootnote}
In addition, we find $\dd\varphi/\dd t_0 = \overline{c}(t_0)$, the
average value of $c$ with distribution $f$ solution of
\begin{equation} \label{dif}
  \frac{1}{2} \, \frac{\partial^2 f}{\partial c^2} +
  v_0\,  \frac{\partial f}{\partial c} +
  (\overline{c} - c) \, f = 0
\end{equation}
with $v_0 \equiv t_0^2-\epsilon _0$. Boundary conditions are 
$(\partial_c f + v_0 \, f)|_{c=0} = 0$ (reflecting wall) and 
$f_0 = f(0)/2$. The diffusion term in (\ref{dif}) reflects the
Gaussian stochastic nature of 2- to 1-clauses reductions, the 
drift term favors small (respectively large) values of the density $c$ 
when $v_0 >0$ (resp. $v_0<0$) --- corresponding to $\delta_2<1$ and 
$\delta _2>1$ respectively --- 
and the third term expresses the relative death-rate 
of search processes with respect to the average rate $\bar c$
(the higher $c$, the more likely it is to encounter
a contradiction)\footnote{Notice that (\ref{dif}) lacks any time derivative 
since $f$ reaches stationarity on a much shorter time-scale
($N^{-1/3})$ than the one relevant for $\Psuccess$ ($N^{-1/6}$).
}. The solution of differential equation (\ref{dif}) reads
$f(c) \propto \exp(- v_0\, c)\; \Ai[\sqrt[3]{2} \,  c + z(v_0)]$, 
where $\Ai$ is the Airy function, $z(x)$ is the inverse function of 
$x(z)= -\sqrt[3]{2} \, \Ai'(z) /\Ai (z)$. Distribution $f$ is shown on
Fig.~\ref{trace_rho} for several values of the drift. Positive
(respectively negative) $v_0$ correspond to trajectories 
below (resp. above) the contradiction line (Fig.~\ref{figure_traj}), 
with distributions $f$  peaked around $c=0$ (resp. $c>0$).

The probability of success remains unchanged (to the leading order in $N$)
once the trajectory exits from the critical region, 
thus $- \ln \Psuccess / N^\expproba = \varphi(t_0\to +\infty)$. This proves
the existence of the scaling function defined in (\ref{fss}), with
\begin{equation} \label{phi} 
\Phi (\epsilon _0) 
= \frac{1}{4} \int_{-\epsilon _0}^{+\infty} 
\frac{\dd x}{\sqrt{\epsilon _0 +x}} \, \big[ x^2 - 2^{2/3}\, z(x)  ]
\end{equation}
Scaling function $\Phi$ is plotted and compared to numerics
in Fig.~\ref{figure_Phi}. It is an easy check that all the results
related to the successful and failure regimes \eg the values
$c^*,\omega^*$ listed above for
finite $\epsilon$ and $N\to \infty$ 
are found back when $\epsilon _0 \to \pm\infty$ respectively.

{\em Universality.} The critical point of R, or any algorithm A that
implements UP \eg GUC \cite{Fri} or HL \cite{kkl} where variables are
chosen to satisfy 2-clauses or according to their occurrences
respectively, is reached when the resolution trajectory is, for some time
$t_\mathrm{A}$, tangent to the $\adeux=1$ line: $\adeux(t_\mathrm{A} +
\Delta t) - 1 = b\, (\Delta t)^n$ with $n \ge 2$ even integer and $b$
determined from the derivatives of the density $c_2$ of 2-clauses at
$t_\mathrm{A}$.  The tangency condition reflects that the creation of new
edges in $G$, from the reduction of 3- into 2-clauses, precisely
compensates the elimination of edges by UP. Although the resolution
trajectory strongly depends on A, the critical behaviour depends on $n,b$
only, and is thus universal.

The value of $n$ is 2 for R, GUC, HL and generic algorithms A. Therefore,
$\expepsilon =\frac 16 ,\expproba=\frac 13$ independently of A; the
scaling function is $\Phi_\mathrm{A}(\epsilon _0) = r_\mathrm{A}^\Phi \;
\Phi (r_\mathrm{A}^\epsilon \; \epsilon _0)$ where the $r_\mathrm{A}$'s
are functions of $b$ \eg $r_\mathrm{GUC}^\Phi \simeq 0.9902$,
$r_\mathrm{GUC}^\epsilon \simeq 1.7182$.
 Fig.~\ref{figure_Phi} illustrates
that GUC and HL fall in this UP universality class. Numerical
investigations suggest that more complex algorithms as KCNFS \cite{dub}
do, too.  This universality class is robust against any change, either
induced by algorithms or present in the input data distribution, in the
degree sequence of the clauses graph G, a consequence of the robustness of
the critical component size distribution $Q(S)$ \cite{rgscaling}.

Higher values of $n (\ge 4)$ are exceptionally found for finely-tuned
input data statistics \eg with clauses of different lengths $\ell
(\le K)$ and appropriate ratios $\alpha _\ell$ (so far we have
restricted to $K=3$ with $\alpha _3 \equiv \alpha)$. If the ratios at
the critical point $\delta _2=1$ are such that the reduction of
$(\ell+1)$- to $\ell$-clauses compensates the disappearance of
$\ell$-clauses for all $2\le\ell < K$, then the resolution trajectory
will stay longer in the critical region, making $\Psuccess$
decrease. The precise condition is $\alpha _\ell =
2^{\ell-1}/\ell/(\ell -1)$ at criticality \cite{gold}, leading to
$\expproba = \frac {n-1}{3n}$ with $n=K-1$~%
\citefootnote{\notegaussienne}.

It would be interesting to extend our study to structured input data
distributions \eg leading to clause graphs G embedded in
finite-dimensional spaces, and possibly to $\expepsilon\ne\frac 13$.
In this context, developing renormalization tools to
capture the critical behaviour of algorithms would be the natural
yet apparently difficult next step. It would also be worth
to study universality for other types of algorithms
\eg local search procedures \cite{extra}, or other
computational tasks \cite{extra2} \eg graph coloring \cite{Moore}, 
where poly/exp transitions take place.

\end{document}